\documentclass[aps,prl,superscriptaddress,preprintnumbers,
showpacs,legalpaper,twoside,twocolumn]{revtex4}

\usepackage{bm}
\usepackage{graphicx}
\usepackage{amsfonts}
\usepackage{amsmath}
\usepackage{amssymb}

\begin{document}

\title{Quantum emulsion: a glassy phase of bosonic mixtures in optical lattices}

\author{Tommaso Roscilde}
\affiliation{Max-Planck-Institut f\"ur Quantenoptik, Hans-Kopfermann-strasse 1,
85748 Garching, Germany}
\author{J. Ignacio Cirac}
\affiliation{Max-Planck-Institut f\"ur Quantenoptik, Hans-Kopfermann-strasse 1,
85748 Garching, Germany}

\begin{abstract}
We numerically investigate mixtures of two interacting bosonic species 
with unequal parameters in one-dimensional optical 
lattices. In large parameter regions full phase segregation is 
seen to minimize the energy of the system, but the true ground 
state is masked by an exponentially large number of metastable
states characterized by \emph{microscopic} phase separation.
The ensemble of these \emph{quantum emulsion} states, reminiscent
of emulsions of immiscible fluids, has macroscopic properties 
analogous to those of a Bose glass, namely a finite compressibility
in absence of superfluidity. Their metastability is probed 
by extensive quantum Monte Carlo simulations generating a rich
correlated stochastic dynamics. The tuning of the repulsion
of one of the two species via a Feshbach resonance drives 
the system through a quantum phase transition to the superfluid 
state. 
\end{abstract}
\maketitle

 Trapped ultracold atoms in optical lattices (OLs)
 offer unprecendented
opportunities of studying the quantum behavior
of correlated matter at a very fundamental level,
namely by experimentally implementing  
Hamiltonians which would have been regarded 
in the past as toy models \cite{Greineretal02}. 
 The introduction of quenchend randomness widely enlarges 
the range of quantum phases which can be explored 
in such systems. In the weakly interacting regime 
the presence of disorder promises to realize
Anderson localization of coherent matter waves
\cite{anderson}; in presence of interaction between 
bosons, condensate fragmentation into macroscopically
many localized states or Anderson localization 
of collective modes leads to the insulating  
\emph{Bose-glass} phase \cite{Fisheretal89}. 
Yet a challenging issue is how to introduce model
disorder in the system. In general all
disorder potentials realizable experimentally show
some form of spatial correlation: 
if the disorder correlation length is much 
larger than the typical correlation length of the 
particles, the disorder imposes a simple density 
modulation fully captured by a local density 
approximation \cite{Schulteetal06},
implying that no new phase is established
in the system. This is the conclusion 
of the seminal experiments carried out
with speckle-laser potentials \cite{Schulteetal06,speckles}.
An alternative way of producing disorder
optically is by loading the atoms in optical 
quasicrystals, for which an insulating behavior 
with unconventional spectral properties has been 
reported recently \cite{Fallanietal06}.

 In this paper we explore an alternative
proposal to introduce disorder in a system
of cold bosons in an OL. A 
second species of $b$ bosons is added to the 
system, having a much smaller hopping amplitude
than the primary $a$ bosons
and therefore producing an effective quasi-static
disorder potential if prepared in random density 
configurations or in a quantum superposition thereof 
\cite{GavishC05,Paredesetal05}. 
The original proposal of Ref. \onlinecite{Paredesetal05}
involved the separate preparation of the 
two bosonic species followed by the sudden onset
of the interspecies interaction, possibly
leading to out-of-equilibrium localization of the 
fast-moving bosons. Here we explore the simpler case 
in which the two species are adiabatically prepared 
together, aiming at their joint ground state 
at $T=0$. We base our analysis on quantum Monte Carlo
simulations of a one-dimensional system, targeting the 
ground state either through \emph{thermal annealing}, 
\emph{i.e.} by slowly cooling the system to $T=0$, and 
through \emph{quantum annealing}, \emph{i.e.} by a slow 
change of the Hamiltonian parameters at low temperature.
Our central finding is that, for weak interactions
between the $a$ bosons, the system displays an exponentially
big number of metastable states which make the 
true ground state of the system essentially 
unreachable. While the ground state would show
perfect phase segregation between the two species,
in the metastable states the two species are
fragmented into small droplets with short-range 
density-density correlations, implying  
phase separation over the length scale of a 
few lattice spacings (depending on the annealing
rate). These \emph{quantum emulsion} states have an 
energy that depends roughly linearly on the 
phase interface, reminiscent of the behavior 
of metastable emulsions of immiscible fluids
\cite{emulsion}. Both upon thermal or quantum
annealing, a significant deviation from the 
true ground state energy is observed 
(\emph{residual energy}), which decays as a 
small power of the annealing rate; this behavior 
is analogous to that of spin 
glasses \cite{Grestetal86, Santoroetal02}. 
The density arrangements of the $b$ bosons in the 
metastable ensemble realize the statistics
of a weakly correlated disordered potential. 
Consistently, the global properties of the $a$ bosons 
in the quantum emulsion states are analogous
to those of a Bose glass, namely gapless
excitations and a finite 
compressibility in absence of superfluidity
\cite{Fisheretal89}.

\begin{figure}[h]
\begin{center}
\includegraphics[
     width=84mm,angle=0]{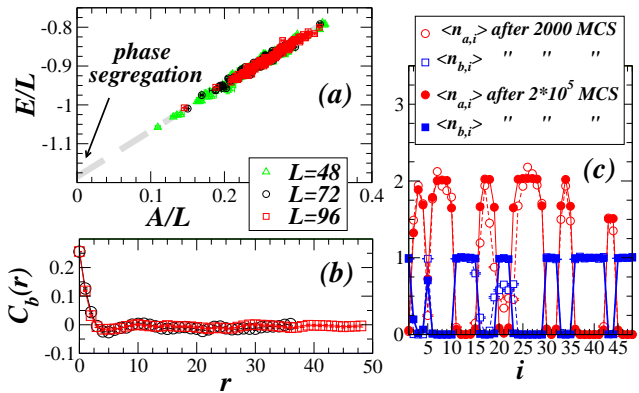} 
\caption{(Color online) $(a)$ Correlation between the energy and 
the phase interface in the metastable ensemble. The model parameters
are $U_a=J_a$, $U_b=U_{ab}=5J_a$, $J_b=J_a/5$, $\langle n_a \rangle = 1$
and $n_b=1/2$. Each point is obtained after 1000 thermalization 
MCS and 2000 measurement MCS after cooling at a rate $r=2*10^{-4}$. 
$(b)$ Density-density correlation function
of the $b$ bosons, 
$C_b(r) = \langle (n_{b,i}-\bar{n}_b) (n_{b,i+r}-\bar{n}_b)\rangle$.
$(c)$ Two density snapshots of a long QMC simulation.}
\label{f.meta}
\end{center}
\vspace*{-1cm}
\end{figure}   

 We describe the bosonic mixture in a
 one-dimensional OL via a
 two-flavor Bose-Hubbard model \cite{Albusetal03}
\begin{eqnarray} 
\null\hspace*{-2cm}{\cal H} = \sum_{i} \Big[
-J_{a}\left(a_{i} a^{\dagger}_{i+1} +{\rm h.c.}\right) 
-J_{b}\left(b_{i} b^{\dagger}_{i+1}+{\rm h.c.}\right)
  \nonumber \\
 + \frac{U_{a}}{2}~ n_{a,i}(n_{a,i}-1)+ 
\frac{U_{b}}{2}~ n_{b,i}(n_{b,i}-1) + 
U_{ab} n_{a,i} n_{b,i}\Big].
\label{e.hamiltonian}
\end{eqnarray}
 
 where $J_{a(b)}$ is the hopping amplitude,
 $U_{a(b)}$ is the on-site intra-species
 repulsion, and $U_{ab}$ is the inter-species
 repulsion. Experimentally, a bosonic mixture of,
 \emph{e.g.}, $^{87}$Rb or of Na
in two different hyperfine states \cite{mixtures}
loaded in spatially anisotropic OLs realizes the 
above Hamiltonian (neglecting the trapping potential).
Due to the different dipolar coupling of the two species
to the OL, the hopping amplitudes $J_a$ and $J_b$ can be 
significantly different from each other. 
In what follows we take the case of slow $b$ bosons, 
$J_b = J_a/5$. The depth of the OL fixes 
the ratios $U_{ab}/J_a = U_{b}/J_a = 5$ 
\cite{footnote}. 
Finally, the application of a magnetic field 
in the proximity of a Feshbach resonance
allows to tune $U_a$. This enables to explore the
various regimes of strong \emph{vs.} weak interaction
for the mobile bosons, which in presence of 
quenched disorder correspond to different
localized phases competing with superfluidity 
\cite{BG}. 
  
 We study the Hamiltonian Eq.~(\ref{e.hamiltonian}) 
making use of the Stochastic Series Expansion 
(SSE) quantum Monte Carlo based on the directed-loop
algorithm \cite{SyljuasenS02} with optimized 
transition probabilities \cite{Polletetal04}.
To efficiently reproduce the correlations between
the two bosonic species we also introduce a 
novel \emph{double-worm} algorithm which allows
for a simultaneous correlated update of worldlines of 
both species \cite{Roscildeprep}. One Monte Carlo step
(MCS) is composed of a diagonal update 
\cite{SyljuasenS02} and as many single- and 
double-loop updates
as to visit on average all SSE vertices twice.
Our simulations are performed in a 
mixed ensemble, namely we fix the density $\bar{n}_b =1/2$
of $b$ bosons \cite{footnote}, 
while a chemical potential term
$-\mu_a \sum_i n_{a,i}$ is added to adjust the
average filling $\langle n_a \rangle$ to a fixed
value. Throughout the paper we choose $\mu_a$ such that
$\langle n_a \rangle \approx 1$ (within less that 2\%).
Chains with periodic boundary conditions and sizes 
up to $L=96$ are investigated, at
inverse temperatures $\beta J_a = L/2$ or higher
(no significant difference is appreciated with
$\beta J_a = L$). We allow for up to 5 (3) $a$ ($b$) 
bosons per site to reproduce the softcore limit. 
 
 A first empirical observation concerns
the difficulty of QMC in equilibrating the
system in the case $U_a \lesssim U_{ab}$,
despite the non-local algorithms used for
the QMC update. This suggests immediately that
many metastable states are present in the system,
and even a non-local stochastic dynamics, 
as the one generated in the QMC update supplied
with multi-cluster moves, is unable to efficiently
escape from local energy minima. At the same time, 
the metastable states are very robust, resisting
essentially unchanged to more than $10^5$ MCS
(see Fig. \ref{f.meta}). 
This is strongly 
reminiscent of glassy systems as, \emph{e.g.},
spin glasses, for which the phase space is essentially 
fragmented into metastable "valleys" which are
not efficiently connected to each other by the dynamics
of the system.
A fundamental ingredient
which defines the class of observed
metastable states is the \emph{annealing} 
procedure followed while preparing the system.
Given that SSE is intrinsically a 
finite-temperature method, we adopt a thermal
annealing scheme by starting at a temperature
$T_{\rm max} = 2J_a$ \cite{Tmaxnote} and linearly 
cooling the system in steps of 
$\Delta T = 0.02 J_a$ down to the physical $T=0$.
At each temperature step we wait $M$ MCS, which
define the \emph{cooling rate} $r$ as $r=\Delta T/M$.
Fig. \ref{f.meta} shows the energy 
$E=\langle {\cal H} \rangle$ of the 
metastable states, reached by repeatedly cooling 
down to $T=\Delta T$ at a rate $r=2\times 10^{-4}$, 
as a function of the \emph{phase interface} 
$A = \sum_i \left\langle \theta(n_{a,i}) \theta(n_{b,i+1})+
\theta(n_{b,i}) \theta(n_{a,i+1}) \right\rangle$
($\theta$=Heaviside function),
giving the average number of nearest neighboring
site pairs having different species on the two sites,
and essentially counting the number of droplets.
There is a clear correlation between energy and
phase interface, indicating that the energy is
minimized for minimal $A$, namely for \emph{phase segregation} 
\cite{Polletetal06, Mishraetal06}.
Yet for all the metastable states the phase 
interface is a significant fraction of the system volume,
which means that these states are microscopic emulsions
of the two bosonic species, with short-ranged
density-density correlations (Fig. \ref{f.meta}). 

It is interesting to observe
that the rescaled data for different system sizes 
all collapse onto the same,
approximately linear dependence of $E$ on $A$, 
whose slope gives the \emph{surface tension}
of the microscopic emulsion. This surface tension
has an exquisite quantum mechanical origin
(hence the name of quantum emulsion). In fact,
while the surface tension in classical emulsions
comes from long-range attractive interactions between 
the atoms/molecules \cite{emulsion}, 
in our case the interactions are 
all repulsive and on-site only. In absence of 
the quantum kinetic term, the energy does not
depend at all on the phase interface, but only on the
phase overlap. Therefore the surface tension stems 
from the attempt to minimize the 
quantum zero-point kinetic energy. 

The almost univocal dependence
of $E$ on $A$ indicates that spatial permutations
of the emulsion droplets leaving the phase
interface unchanged produce quasi-degenerate
states. The collapse of the data for different
sizes shows moreover that the typical droplet
size is only weakly dependent
on the system size. Therefore the number of 
droplets in the emulsion scales linearly with the 
system size, and so the number of droplet permutations
scales \emph{exponentially}. This implies that the 
number of quasi-degenerate emulsion states 
scales also exponentially. 
Hence the strong tendency of the system to 
get trapped in the metastable states.

\begin{figure}[h]
\begin{center}
\includegraphics[
     width=83mm,angle=0]{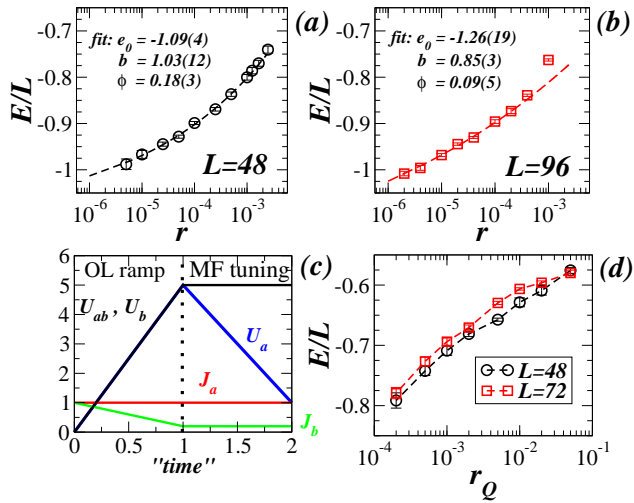} 
\caption{(Color online) ($a$-$b$) Energy $vs.$ cooling rate 
for thermal annealing (Hamiltonian parameters as in Fig. \ref{f.meta}).
$\Delta T = 0.05 J_a$ for $L=48$ and $0.02 J_a$ for $L=96$.
($c$) Time schedule in the quantum annealing procedure;
in the simulation the time variable is changed in steps 
of $5*10^{-2}$. MF=magnetic field. ($d$) 
Energy $vs.$ annealing rate  
for quantum annealing.}
\label{f.crate}
\end{center}
\vspace*{-0.4cm}
\end{figure}

 Fig. \ref{f.crate} shows the dependence of 
the average energy of the metastable ensemble 
at $T=\Delta T$ (see Fig. caption) as a function of the cooling rate $r$. 
Averaging over repeated cooling cycles is made,
with $\sim 200$ takes for the fastest cooling rates 
and $\sim 20$ takes for the slowest ones.
It is clear that the final energy systematically
deviates from the true equilibrium value and it
depends on $r$, although the dependence is extremely
slow. A fit to a power-law dependence of the 
kind $E(r)/L = e_0 + b~r^{\phi}$ gives $\phi =0.18(3)$
for $L=48$ and $\phi=0.09(5)$ for $L=96$;
this might suggest that $\phi\to 0$ for
$L\to\infty$, leaving out a logarithmic
dependence only which is generally expected
in presence of low-energy metastable states  
\cite{HuseF86,Santoroetal02}. 
A similar slow dependence on $r$ is also observed
for the average phase interface, as expected
from the linear relation with the energy.
Therefore the kinds of quantum emulsion reached 
at different cooling rates are weakly dependent on 
$r$ itself, and the behavior described before
for $r=2*10^{-4}$ is generic for a broad range of 
$r$ values. 

 Although thermal annealing is quite convenient
for QMC simulations, it is not necessarily the 
most relevant annealing procedure for experiments.
In fact, a typical OL experiment
is in principle performed at fixed, very low temperature, 
changing the Hamiltonian of the
bosons from that of a trapped weakly interacting gas
to that of a strongly interacting one in the OL. 
Therefore the picture of so-called
\emph{quantum} annealing \cite{Santoroetal02,Dasetal05} 
is more appropriate for the experimental preparation 
of the system. We simulate an \emph{incoherent} two-step 
quantum annealing process in which first
the ratios $J_{a}/U_{a(b)}$ and $J_b/J_a$ 
are linearly decreased, mantaining 
$U_a = U_b = U_{ab}$ (OL ramp-up), and
then $U_a$ is decreased by tuning an applied
magnetic field close to a Feshbach resonance. 
The whole annealing is performed with 
$M$ MCS, defining the annealing rate 
$r_{\rm Q} = 1/M$. We observe that also upon 
quantum annealing the system remains far from the true
equilibrium; we cannot fit $E$ $vs.$ $r_{\rm Q}$
neither with a power law (as in the thermal case) 
nor with a logarithm, which suggests
that even for the slowest annealing rates 
we are far from the asymptotic behavior 
for $r_{\rm Q}\to 0$.

 Having shown the robust features of 
the ensemble of metastable emulsion states,
we argue that this ensemble is experimentally
relevant. The existence of exponentially many
metastable states implies the presence
of many avoided level crossings 
between such states upon an arbitrary evolution 
of the Hamiltonian parameters
toward their final value. The gaps
at the avoided level crossings are exponentially
small in the ratios $J_{a(b)}/U_{a(b)}$,
given that the metastable states typically differ 
from each other by a macroscopic rearrangement 
of the density distribution. The evolution of the
system's state during the experimental 
transformation of the Hamiltonian parameters will
then exhibit a cascade of Landau-Zener (LZ) 
processes, so that the final state $|\Psi\rangle$ of the system will 
be a large quantum superposition  $|\Psi\rangle = \sum_\psi c_\psi
|\psi\rangle$ of different
metastable states $|\psi\rangle$ \cite{footnote2}.


\begin{figure}[h]
\begin{center}
\includegraphics[
     width=65mm,angle=0]{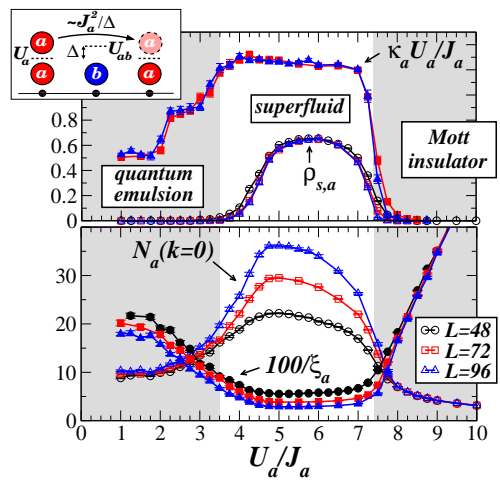} 
\caption{(Color online) Phases of the $a$ bosons
upon tuning the repulsion $U_a$. Upper panel:
superfluid fraction $\rho_{s,a}$ and 
compressibility $\kappa_a$. Lower panel:
coherent fraction $N_a(k=0)$ and inverse
correlation length $\xi_a^{-1}$. In the inset:
sketch of the interaction-assisted resonant
tunneling.}
\label{f.Uscan}
\end{center}
\vspace*{-0.6cm}
\end{figure} 

 The above argument motivates us to regard the
metastable emulsion states as the experimentally
relevant ensemble for a given annealing procedure:
incoherent averaging over this ensemble 
essentially corresponds to measurements in the
experimental final state of the system \cite{footnote2}. 
The averaging over the emulsion states reveals an 
out-of-equilibrium \emph{quantum emulsion phase},
with exotic macroscopic properties. Fig.~\ref{f.Uscan}
shows the global properties of the $a$ bosons
as a function of the tunable $U_a/J_a$ repulsion upon
averaging over 100-200 thermal annealing cycles 
at a rate $r=2*10^{-4}$. It is observed that
for $U_a/J_a \lesssim 3.5$ the $a$ bosons are in a
quantum emulsion phase with zero superfluid
fraction $\rho_{s,a}$ (estimated through winding
number fluctuations \cite{SyljuasenS02,PollockC87})
but finite global compressibility 
$\kappa_a = \beta\langle [\sum_i (n_{a,i} - 
\langle n_a \rangle)]^2 \rangle /L$. 
The absence of superfluidity is clearly
due to the fragmentation of the $a$ bosons into
incoherent metastable droplets. Such fragmentation
can be regarded effectively as a phenomenon
of localization due to the strong repulsion
with the $b$ bosons, which are also fragmented
into weakly correlated droplets. In this respect,
the quantum emulsion phase behaves as 
a metastable Bose glass: although 
not superfluid, this phase is compressible
because arbitrarily large droplets of $a$ bosons 
can appear which admit gapless particle number
fluctuations. 

 Upon increasing the on-site repulsion $U_a/J_a$,
the localization of $a$ bosons can be overcome by 
a phenomenon of interaction-assisted resonant tunneling.
An effective hopping $\approx J_a^2/(U_{ab}-U_a)$ 
(to 2$^{\rm nd}$ order perturbation theory)
brings an $a$ boson from a doubly occupied site 
to a singly occupied one across an intermediate site 
occupied by a $b$ boson (see Fig.~\ref{f.Uscan}).
Close to the resonance condition $U_a \approx U_{ab}$
this process becomes very effective, and the coherent
tunneling of $a$ bosons from one droplet to the 
neighboring one leads to the onset of superfluidity.
Roughly speaking, when $n_b \ll 1$ a fraction 1-$n_b$ of 
$a$ bosons screens the potential created by the $b$-boson 
droplets, and it assists the resonant tunneling of
the remaining fraction $n_b$, which then leads to 
$\rho_{s,a} \sim n_b$. The phase of the system
with superfluidity of the $a$ bosons still shows 
a significant metastable behavior: the spatial arrangement
of the $b$-boson droplets is in fact quasi-static in the
QMC simulation, namely the simulation is strongly 
non-ergodic in that translational invariance remains
broken over more than $10^5$ MCS. It is suggestive 
to picture the $b$ bosons and the normal fraction of 
the $a$ bosons as frozen in a metastable quantum
emulsion, whereas the superfluid fraction of $a$
bosons is able to coherently tunnel through the 
emulsion and to ergodically visit the entire lattice.
Concerning experimentally relevant quantities, the onset 
of superfluidity is marked by a dramatic enhancement 
of the coherence peak 
$N_a(k=0)= \sum_{ij} \langle a_i^{\dagger} a_j \rangle /L$,
and by a strong decrease of its width (quantified 
by the inverse phase correlation length $\xi_a^{-1}$ 
through second-moment estimation).                        
Upon further increasing $U_a/J_a$ the unit-filled
$a$ bosons are eventually driven to a Mott insulating
state, forming an almost uniform background
for the $b$ bosons which delocalize into a 
superfluid state.

 In conclusion, we have reported a strongly metastable
 behavior of repulsive bosonic mixtures in one-dimensional 
 optical lattices. An out-of-equilibrium quantum emulsion 
 phase emerges from QMC simulations, and it is found
 to be very robust to thermal as well as quantum annealing.
 This phase has a pronounced glassy behavior, with
 a significant residual energy slowly decaying 
 with the annealing rate. All these ingredients
 suggest that this phase be extremely relevant 
 for ongoing experiments on bosonic mixtures
 as well as for one-dimensional fermion-boson mixtures 
 in optical lattices. 
 
     

   Useful discussions with S. D\"urr, G. Rempe, N. Syassen,
   O. Sylju\aa sen, and T. Volz are gratefully acknowledged.
   This work is supported by the E.U. through the 
   SCALA integrated project.


\begin{thebibliography}{99}
\bibitem{footnote} A broader range of the Hamiltonian parameters
has been investigated \cite{Roscildeprep}, and the behavior 
reported here is found to be quite generic.
\bibitem{Tmaxnote} No significant difference is 
observed if starting from higher temperatures.
\bibitem{footnote2} Even in the ideal case of a fully coherent evolution, 
the expectation values of local observables
over the final state $\langle \Psi |(...)|\Psi \rangle$ should be 
well approximated by an incoherent average over the metastable 
ensemble, $\sum_\psi |c_\psi|^2 \langle \psi|(...)|\psi\rangle$, 
given that the various $|\psi\rangle$ components generally differ 
by a macroscopic density rearrangement of the two species.
\end{thebibliography}
\end{document}